%% file: article_4.tex
\documentclass[
10pt, % Main document font size
a4paper, % Paper type, use 'letterpaper' for US Letter paper
oneside, % One page layout (no page indentation)
headinclude,footinclude, % Extra spacing for the header and footer
BCOR5mm, % Binding correction
]{scrartcl}

\input{structure.tex} % Include the structure.tex file which specified the document structure and layout

\title{\normalfont\spacedallcaps{Election of government ministers}} % The article title

\author{Itai Lashover\textsuperscript{1} \\ \href{mailto:itai.lash@gmail.com}{itai.lash@gmail.com} 
   \and Liav Weiss\textsuperscript{1} \\ \href{mailto:liavweiss@gmail.com}{liavweiss@gmail.com} 
   \and Amichai Kafka\textsuperscript{1} \\ \href{mailto:amichaikp@gmail.com}{amichaikp@gmail.com}
   \and Shoshana Levin\textsuperscript{1} \\ \href{mailto:shani032@gmail.com}{shani032@gmail.com}
   \and  {\small Instructor: Erel Segal Halevi}}

\date{July 24, 2022} % An optional date to appear under the author(s)

\begin{document}

%----------------------------------------------------------------------------------------
%	HEADERS
%----------------------------------------------------------------------------------------

\renewcommand{\sectionmark}[1]{\markright{\spacedlowsmallcaps{#1}}} % The header for all pages (oneside) or for even pages (twoside)
\lehead{\mbox{\llap{\small\thepage\kern1em\color{halfgray} \vline}\color{halfgray}\hspace{0.5em}\rightmark\hfil}} % The header style
\pagestyle{scrheadings} % Enable the headers specified in this block

%----------------------------------------------------------------------------------------
%	TABLE OF CONTENTS
%----------------------------------------------------------------------------------------

\maketitle % Print the title/author/date block

\setcounter{tocdepth}{2} % Set the depth of the table of contents to show sections and subsections only

\tableofcontents % Print the table of contents

%----------------------------------------------------------------------------------------
%	AUTHOR AFFILIATIONS
%----------------------------------------------------------------------------------------

\let\thefootnote\relax\footnotetext{\textsuperscript{1} \textit{Computer science students, Ariel University, Israel.}}

%----------------------------------------------------------------------------------------
%	ABSTRACT
%----------------------------------------------------------------------------------------

\section*{Abstract} % This section will not appear in the table of contents due to the star (\section*)

The executive branch (the government) is usually not directly elected by the people, but is created by another elected body or person such as the parliament or the president.
As a result, its members are not directly accountable to the people, individually or as a group.

We propose a scenario where government members are directly elected by the people, and seek to achieve proportional representation in the process.

We will present a formal model for the allocation of K offices, each associated with a disjoint set of candidates contesting for that seat.

A group of voters provides ballots for each of the offices.
Since using simple majority voting for each office independently may result in minority preferences being completely ignored, here we adapt the greedy version of proportional approval voting (GreedyPAV) to our framework.

In the article Electing the Executive Branch\cite{Electing:2021} you can find an in-depth explanation of the model and a demonstration - through computer-based simulations - of how voting for all offices together using this rule overcomes this weakness and upholds the axiom of proportionality.

In this article, we will present the implementation of the algorithm (GreedyPAV) proposed by Rutvik Page, Ehud Shapiro, and Nimrod Talmon in the article mentioned above\cite{Electing:2021}.
In addition, we tested our implementation through a survey, the results of which will be presented and analyzed later in the article.

%----------------------------------------------------------------------------------------
%	INTRODUCTION
%----------------------------------------------------------------------------------------

\section{Introduction}

Consider a scenario in which a government in a country has to be populated;
i.e., there should be elected members of the government like the minister for health, the minister for education, etc. Usually this assignment process is done via a non-participatory process .
In this paper we will detail the problem and how we implemented the solution mentioned in the article\cite{Electing:2021}, by building a website that makes the solution accessible to voters.
We will present a survey that simulates the scenario mentioned above and analyze its results.
 
%----------------------------------------------------------------------------------------
%.    THE PROBLEM WITH THE CURRENT ELECTION SYSTEM 
%----------------------------------------------------------------------------------------

\section{The problem with the current election system}

In most democratic countries, the selection of ministers in the government is done through coalition agreements or by the prime minister/president.
As a result, the voter has no direct influence on the formation of the government and the choice of ministers, and the government ministers are elected in a way that does not always reflect the will of the voters.

%----------------------------------------------------------------------------------------
%	METHOD
%----------------------------------------------------------------------------------------

\section{Method}
In this section, we will describe the way we chose to realize the solution to the problem mentioned in the previous section.
We used the GreedyPAV algorithm\cite{Electing:2021} and set up a website\cite{ourGovernment} that makes the use of the algorithm accessible to voters.

%------------------------------------------------
%	GREEDYPAV
%------------------------------------------------
\subsection{GreedyPAV}

The GreedyPAV voting rule devised and published by Rotvik Page, Ehud Shapiro and Nimrod Talmon in 2021 is used for multi-winner elections and is known to be proportional to that framework.
The GreedyPAV algorithm maintains the ‘Global Justified Representation’ definition which means for each subset of voters of a certain size that agrees on at least one candidate there is a selected candidate that they agree on.

This algorithm receives as input the list of offices to be occupied, a list of candidates for each of the offices and the list of votes of each of the voters.

The algorithm gives weight to each voter and when the voter is satisfied its weight is reduced. Which ensures that in the next round, the weight of the voters who have not yet been satisfied will be higher.

Finally, the algorithm emits the selected candidate for each of the offices.

%------------------------------------------------
%	OURGOVERNMENT
%------------------------------------------------
\subsection{ourGovernment}

Our Government is a web application that makes the use of the Greedy PAV algorithm accessible to voters.
Our web application contains two main features:
\begin{itemize}
\item Using the proposed election system that supports a limited number of voters (demo).
The user enters the names of the offices and the names of the candidates for each of the offices, then the user enters the preferences of the voters and finally the algorithm returns a table showing the candidates selected for each of the offices.
In addition, the user is provided with an in-depth explanation of the proportionality and correctness of the allocation.
\item Using the algorithm for an unlimited number of voters by uploading a file containing all the offices, candidates and citizens' votes.
\end{itemize}

\begin{figure}[h]
\centering
\subfloat[OurGovernment HomePage.]{\includegraphics[width=.45\columnwidth, height=.3\columnwidth]{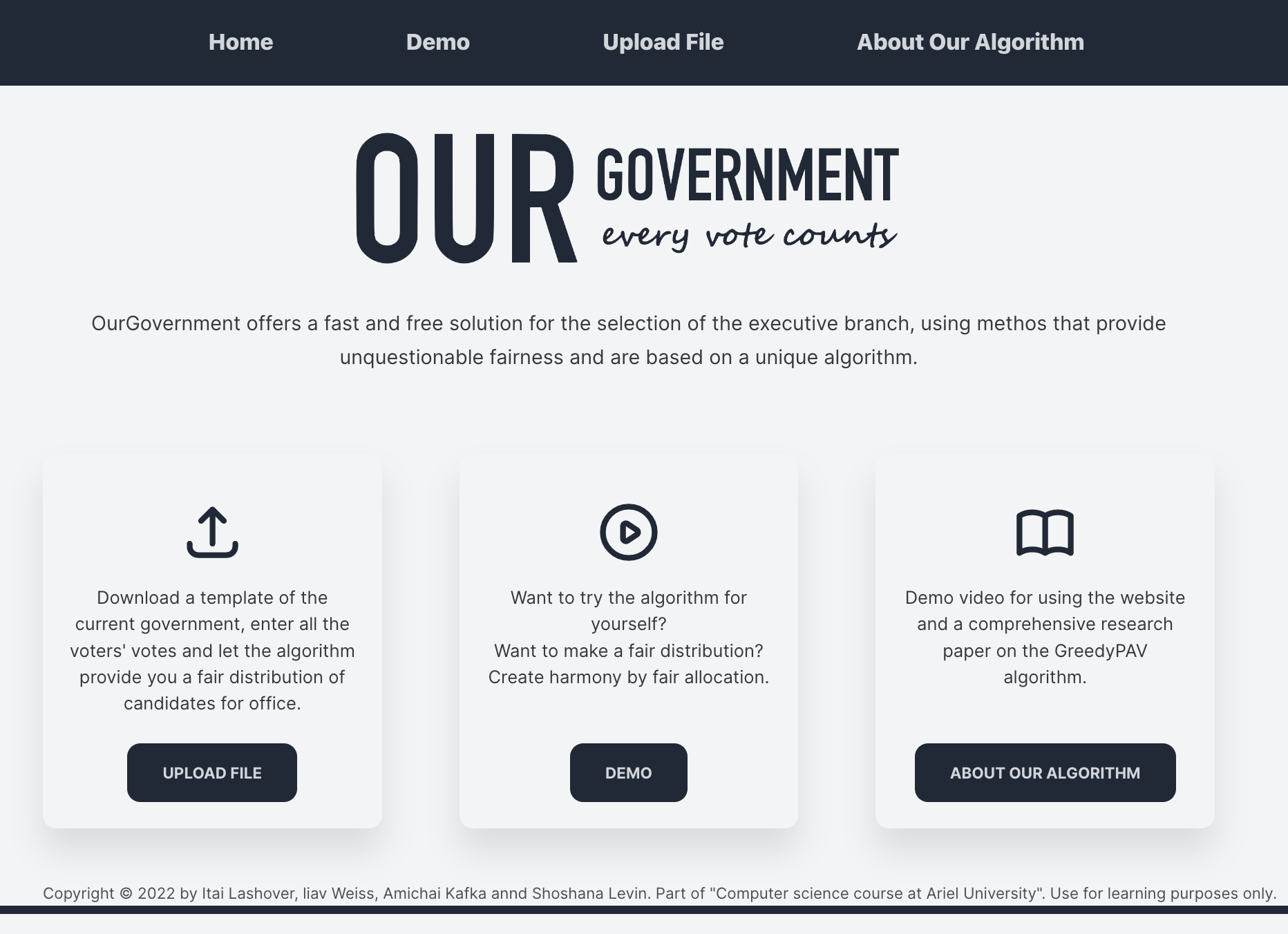}} \\
\subfloat[Demo: Entering offices and candidates.]{\includegraphics[width=.45\columnwidth, height=.3\columnwidth]{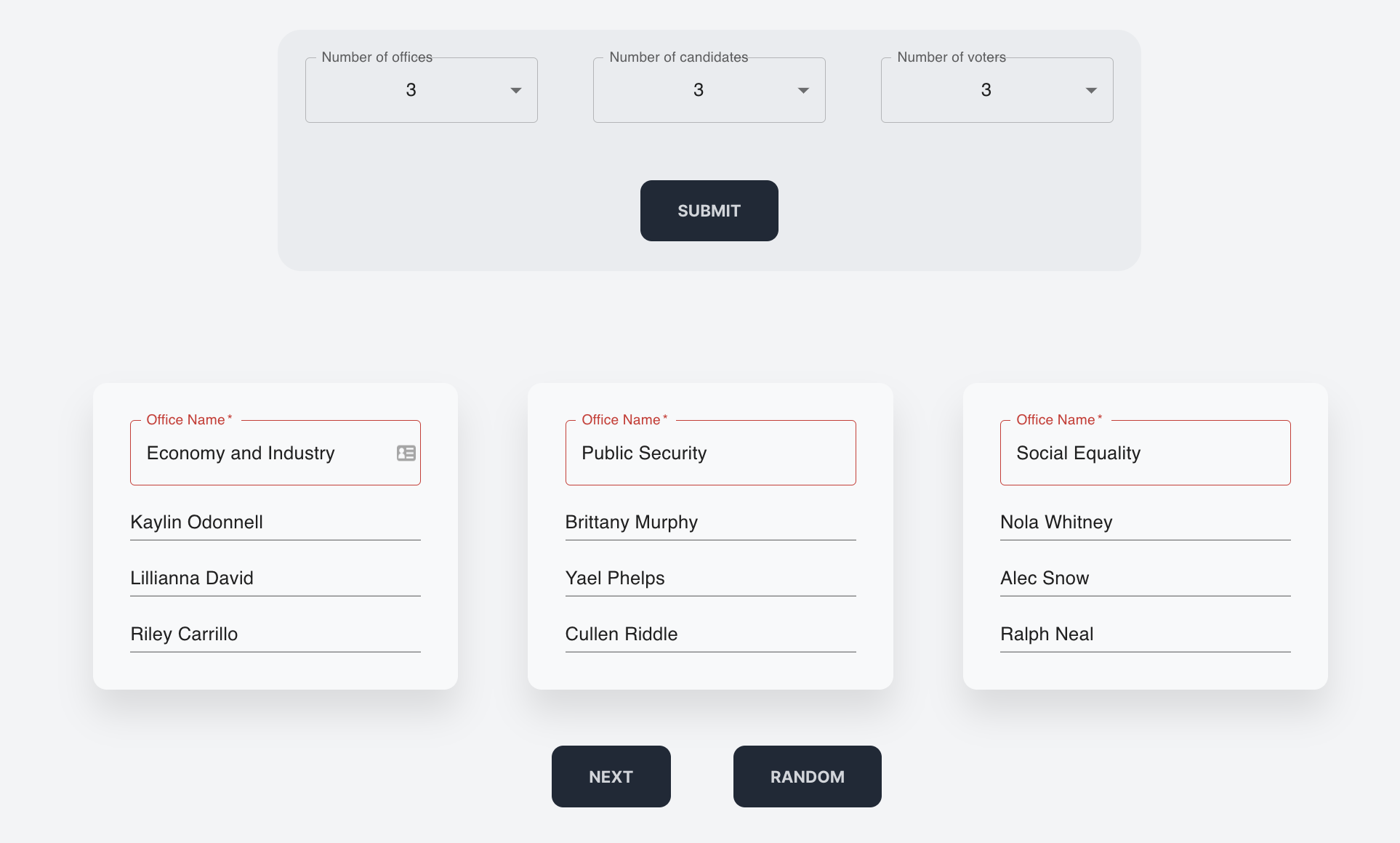}} \quad 
\subfloat[Demo: Entering votes.]{\includegraphics[width=.45\columnwidth, height=.3\columnwidth]{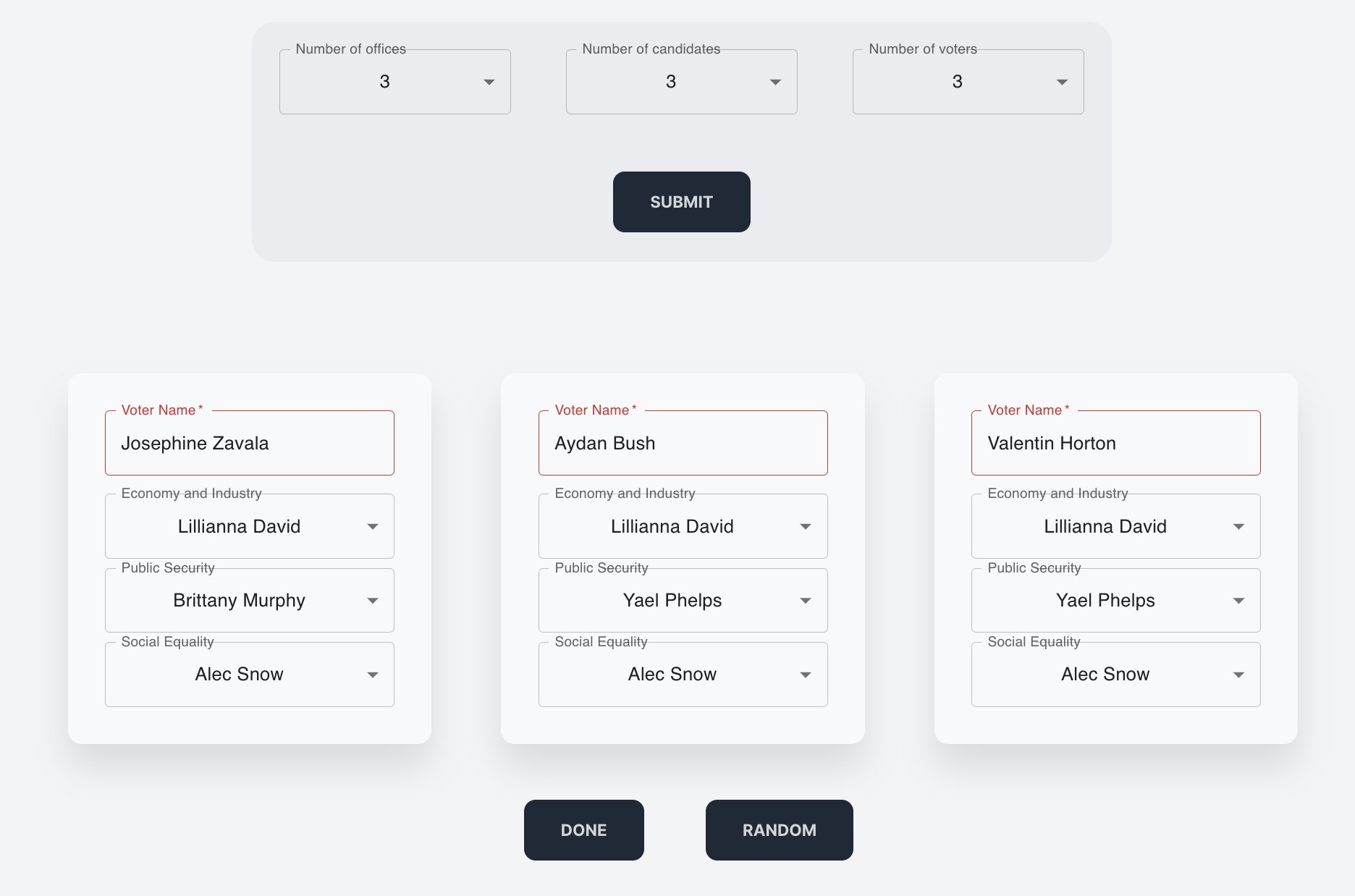}} \\
\subfloat[Demo: Algorithm results.]{\includegraphics[width=.45\columnwidth, height=.3\columnwidth]{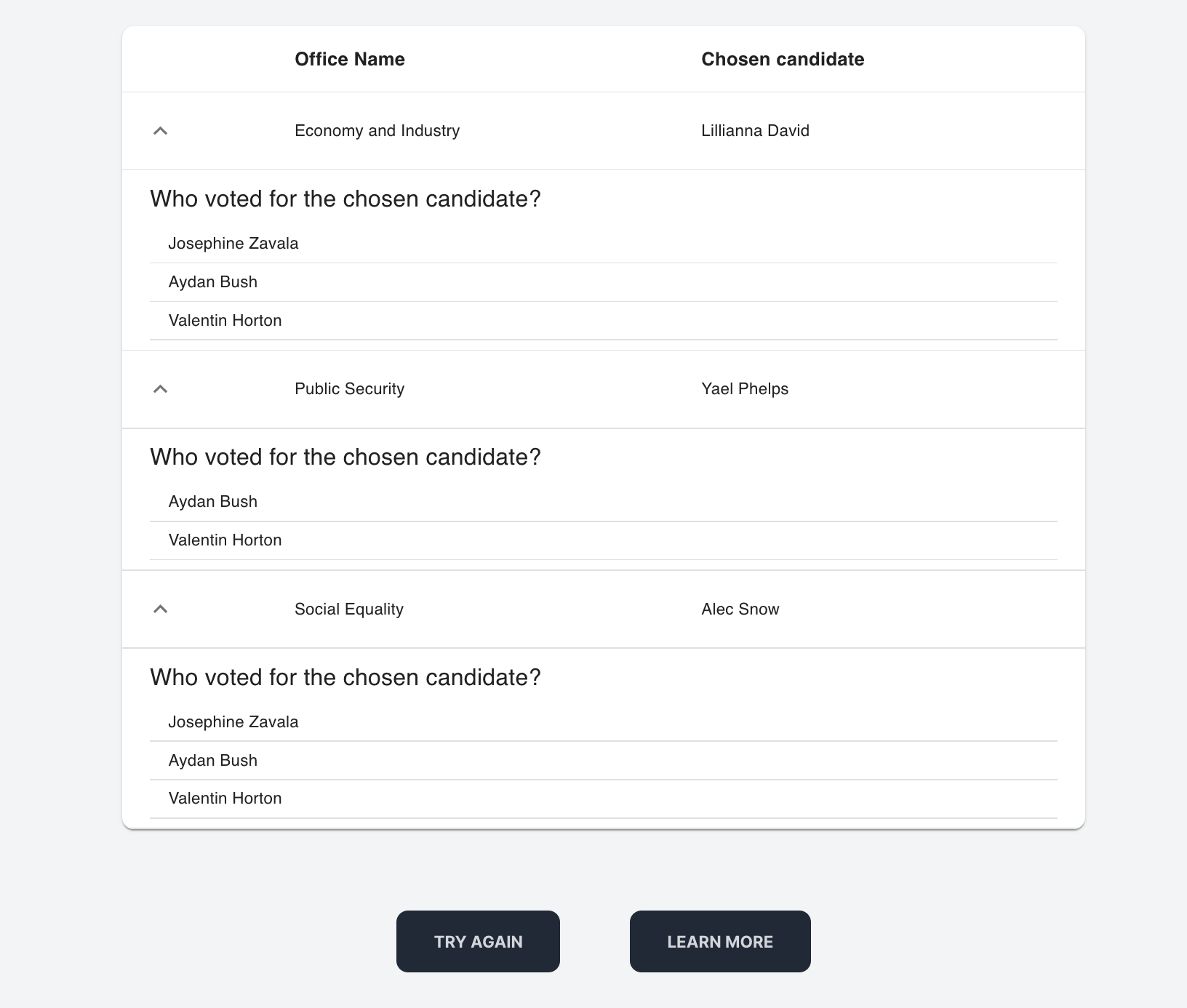}} \quad
\subfloat[UploadFile: The steps to be followed to upload the file.]{\includegraphics[width=.45\columnwidth, height=.3\columnwidth]{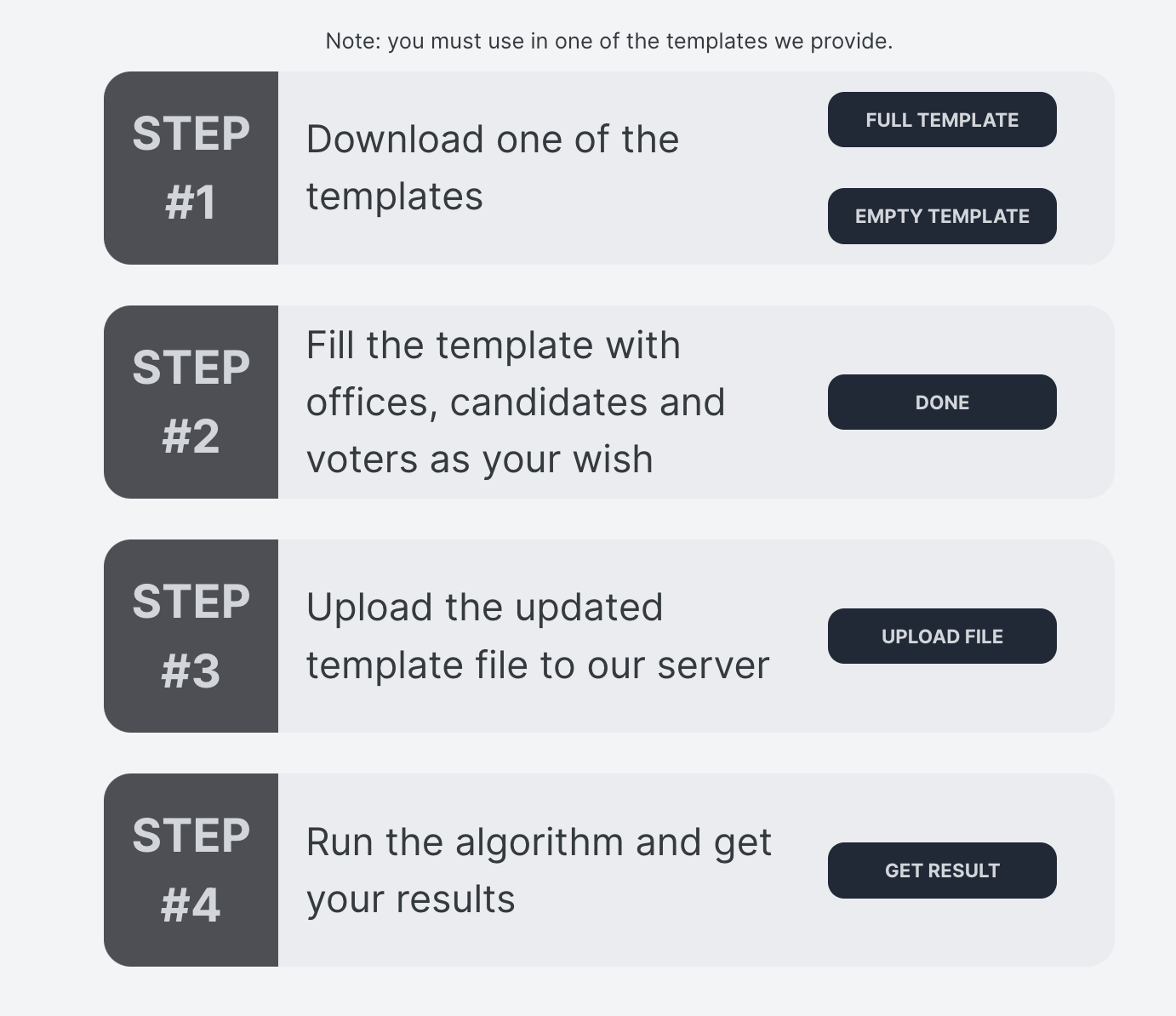}}

\caption[Some pictures from OurGovernment web application.]{Some pictures from OurGovernment web application.} % The text in the square bracket is the caption for the list of figures while the text in the curly brackets is the figure caption
\label{fig:esempio}
\end{figure}

%----------------------------------------------------------------------------------------
%	SURVEY
%----------------------------------------------------------------------------------------

\section{Survey}

In the previous sections, we discussed the importance of the algorithm and its implementation, now, we would like to make sure that the algorithm does solve the problem mentioned on the previous page.
To do this, we conducted a survey with the help of an internet panel of the survey company panel4all for a representative sample of 500 citizens who have the right to vote in Israel.
In the next sections we will present the survey we conducted, its results and their analysis.

%------------------------------------------------
%	INTRODUCING THE SURVEY
%------------------------------------------------

\subsection{Introducing the survey}

This survey proposes a system for selecting ministers in the government, while implementing an innovative algorithm that ensures that the ministers represent the entire public.
The purpose of the survey is to check how well the results of the algorithm proportionally and fairly represent the preferences of the voters.
Each participant was asked to answer the questions according to what he would do if the elections were held today, and in the manner that best reflects his opinion today.
The full survey can be found here\cite{survey}, under the `Survey proposal` pdf file.

All participants may not answer all or part of the questions in the survey.

The survey consists of two parts:

\subsubsection{First part: Choosing preferences for each office.}

In this part, two questions will be presented to each survey participant.
\begin{itemize}
\item In the \textbf{first question} we present the problem, describe to the participant the current election system in the State of Israel (coalition elections) which do not directly represent all voters.
We then present the alternative we propose to the current system in which members of the government are directly elected by the people, and promise to achieve proportional representation of all
the voters as part of the elected government.
After that, each participant was asked whether the alternative we offer is better than the method used today?

\item In the \textbf{second question}, we present to each of the participants a list of 12 ministries in the government where each ministry has four candidates.

We note that the selection of candidates for each of the offices was done by us and does not necessarily reflect the wishes of the candidates.

The list of candidates for each office was made in consultation and cooperation with politicians and members of regional councils.
The current ministers of the 12 ministries were among the candidates, and the three additional candidates for each of the ministries were chosen (more or less) according to their field of interest, their political and public past and their main activities in recent years.

The survey was conducted in the run-up to an election period in the country, so we compiled the lists of candidates in a proportional way according to the estimated number of mandates each party received (according to the most updated election surveys as of the day the survey was sent).
That is, suppose there is an estimate that one of the parties will receive 30 mandates (which is a quarter of the number of mandates in Israel) then 12 of the 48 candidates for the 12 ministries will be members of this party.

Each participant in the survey must choose his preferred candidate for each of the offices.
\end{itemize}

After the 500 participants answered the first part of the survey, the results reach us (the personal identity of the participants is discreet), we enter the preferences of all participants into the algorithm and receive the final results.

\subsubsection{Second part: presenting the results of the algorithm to the participants.}

In this section, two additional questions are presented to the participants.
\begin{itemize}
\item In the first question, we present to them the distribution of the elected ministers to each of the ministries according to the algorithm and point out to them that this distribution is fair and proportional. That is, each group in the population receives proportional representation according to the size of the group.
Should each participant in the survey indicate how satisfied he is with the distribution received.

\item In the second question, according to the results of the algorithm, each participant was asked whether the method for allocating government ministers that we propose is superior to the current system.
\end{itemize}

%------------------------------------------------
%	SURVEY RESULTS
%------------------------------------------------
\subsection{Survey results}
The full survey results can be found here \cite{survey}.

\subsubsection{First part: Choosing preferences for each office.}
Below are the distribution of answers to the second question in the first part:

We asked whether the alternative we offer is better than the election method used today.
\begin{itemize}
\item 22\% of the respondents to the questionnaire answered that they prefer the current election system.
\item 58\% of the respondents answered that they prefer the proposed alternative.
\item The remaining 20\% answered that for them there is no difference between the methods.
\end{itemize}
We will emphasize and note that we asked this question before presenting the results to the respondents.
\subsection{Analysis of survey results}

\subsubsection{Second part: presenting the results of the algorithm to the participants.}
Below is the distribution of the answers to the two questions asked in the second part:

In the first question, the results of the algorithm were presented to the respondents and they were asked how satisfied they were with the results.
\begin{itemize}
\item About 33\% of those surveyed answered that they were satisfied with the results,  24\% of the respondents were indifferent with the results and the remaining 43\% were not satisfied with the results.
\end{itemize}

In the second question, we asked again, whether the alternative we offer is better than the election method used today.
\begin{itemize}
\item 25\% of the respondents to the questionnaire answered that they prefer the current election system.
\item 54\% of the respondents answered that they prefer the proposed alternative.
\item The remaining 21\% answered that for them there is no difference between the methods.
\end{itemize}

%------------------------------------------------
%	SURVEY ANALYSIS
%------------------------------------------------
\subsection{Analysis of survey results}
The full survey analysis can be found here \cite{survey}.

As part of the analysis of the results, we divided the respondents into a minority group and a majority group, represented by religion.
\begin{itemize}
\item The majority group (Jews): 81.7\% of respondents.
\item The minority group (the rest): 19.3\% of respondents.
\end{itemize}
Among the 58\% of those surveyed in the majority group, 37\% were satisfied and the other 21\% were indifferent with the results and among the 55\% of those surveyed from the minority group, 18\% were satisfied and the other 37\% were indifferent with the results.

As you can see, despite the division of those surveyed into two different groups whose opinions are different and the members of the government who represent them are different, their satisfaction with the results is similar.
This indicates a fair distribution because both groups have the same partial representation in the division of ministries to their representatives in the government while we would expect one of the groups that are 'ideological rivals' to be disappointed with the final division of ministries.

It is interesting to note in the analysis above, that although most of the voters are not satisfied/indifferent with the results, most of the voters (54\% of them) still prefer the election system we proposed.

In addition, voters were given the opportunity to justify their satisfaction with the results of the algorithm in free text.
During the analysis of the results, we went through all the answers of the respondents and noticed a phenomenon, a significant part of the respondents who indicated that they were not satisfied with the results of the algorithm added and wrote that they were not satisfied with the choice of a minister or two who represent opinions opposite to their own and therefore indicated that they were not at all satisfied with the final result.

for example:
\begin{itemize}
\item"I have a problem with one choice that I really didn't connect with; But apart from that, I really liked it and it's very true and accurate" 

(answered that is not at all satisfied with the results of the algorithm)
\item"Don't want Anonymous in the government" 

(answered that is not at all satisfied with the results of the algorithm)
\end{itemize}

We note that the central idea of a fair and proportional distribution in our case means that there will certainly be representatives in the government who represent the entire range of different ideologies in society.
Therefore, it is clear that the majority of citizens will not be satisfied with all the elected ministers, but the algorithm guarantees satisfaction with the fairness and proportionality that the algorithm guarantees.

From the analysis of our results, it appears that many of the respondents who answered that they were not satisfied with the results in general were not satisfied with one or two selected representatives, while this situation is desirable and logical in a proportional distribution.

%----------------------------------------------------------------------------------------
%	CONCLUSION
%----------------------------------------------------------------------------------------

\section{Conclusion}
In this paper we presented the implementation of an innovative method for choosing the government, by the greedyPAV algorithm that allows a fair distribution of the votes.
We did a survey using the method and saw that most of the respondents were satisfied/indifferent with the results and finally we saw a majority that prefer the above method over the existing method.

%----------------------------------------------------------------------------------------
%	BIBLIOGRAPHY
%----------------------------------------------------------------------------------------

\renewcommand{\refname}{\spacedlowsmallcaps{References}} % For modifying the bibliography heading

\bibliographystyle{unsrt}

\bibliography{sample} % The file containing the bibliography

%----------------------------------------------------------------------------------------

\end{document}

%% file: structure.tex
\usepackage[
nochapters, % Turn off chapters since this is an article        
beramono, % Use the Bera Mono font for monospaced text (\texttt)
eulermath,% Use the Euler font for mathematics
pdfspacing, % Makes use of pdftex’ letter spacing capabilities via the microtype package
dottedtoc % Dotted lines leading to the page numbers in the table of contents
]{classicthesis} % The layout is based on the Classic Thesis style

\usepackage{arsclassica} % Modifies the Classic Thesis package

\usepackage[T1]{fontenc} % Use 8-bit encoding that has 256 glyphs

\usepackage[utf8]{inputenc} % Required for including letters with accents

\usepackage{graphicx} % Required for including images
\graphicspath{{Figures/}} % Set the default folder for images

\usepackage{enumitem} % Required for manipulating the whitespace between and within lists

\usepackage{lipsum} % Used for inserting dummy 'Lorem ipsum' text into the template

\usepackage{subfig} % Required for creating figures with multiple parts (subfigures)

\usepackage{amsmath,amssymb,amsthm} % For including math equations, theorems, symbols, etc

\usepackage{varioref} % More descriptive referencing

\usepackage{pdfpages}

\theoremstyle{definition} % Define theorem styles here based on the definition style (used for definitions and examples)

\theoremstyle{plain} % Define theorem styles here based on the plain style (used for theorems, lemmas, propositions)

\theoremstyle{remark} % Define theorem styles here based on the remark style (used for remarks and notes)

\hypersetup{
colorlinks=true, breaklinks=true, bookmarks=true,bookmarksnumbered,
urlcolor=webbrown, linkcolor=RoyalBlue, citecolor=webgreen, % Link colors
pdftitle={}, % PDF title
pdfauthor={\textcopyright}, % PDF Author
pdfsubject={}, % PDF Subject
pdfkeywords={}, % PDF Keywords
pdfcreator={pdfLaTeX}, % PDF Creator
pdfproducer={LaTeX with hyperref and ClassicThesis} % PDF producer
}